\documentclass[aps,prl,twocolumn]{revtex4-1}
\usepackage{amsmath,bm,epsfig,,subfigure}
\usepackage{color}
\usepackage[normalem]{ulem}

\def\Fbox#1{\vskip1ex\hbox to 8.5cm{\hfil\fboxsep0.3cm\fbox{%
  \parbox{8.0cm}{#1}}\hfil}\vskip1ex\noindent}  %%  {TEXT} in BOX

\newcommand{\B}[1]{{\bm{#1}}}%% Bold Roman & Greek Lower & Upper Case
\newcommand{\C}[1]{{\mathcal{#1}}}    %%   Calligrapfic Upper case
\let \= \equiv \let\*\cdot \let\~\widetilde \let\-\overline

\begin{document}
\title{Atomistic theory of the shear band direction in amorphous solids}
\author{Ashwin J.$^1$, Oleg Gendelman$^2$,  Itamar Procaccia$^1$ and Carmel Shor$^1$}
\affiliation{$^1$Department of Chemical Physics, The Weizmann
 Institute of Science, Rehovot 76100, Israel\\ $^2$ Faculty of Mechanical Engineering, Technion, Haifa 32000, Israel.}
\date{\today}
\begin{abstract}
One of the major theoretical riddles in shear banding instabilities is the angle that the shear band chooses
spontaneously with respect to the principal stress axis. Here we employ our recent atomistic theory to compute
analytically the angle  in terms of the characteristics of the Eshelby inclusion that models faithfully
the eigenfunction of the Hessian matrix that goes soft at the plastic instability. We show that loading protocols
that do not conserve volume result in shear bands at angles different from 45$^o$ to the strain axis; only when
the external strains preserve volume like in pure shear, the shear bands align precisely at 45$^o$ to the strain axis. We compute an analytic formula for the angle of the shear band in terms of the characteristics of the loading protocol; quantitative agreement with computer simulations is demonstrated.
\end{abstract}

\maketitle

\begin{figure}[h]
\includegraphics[scale = 0.28]{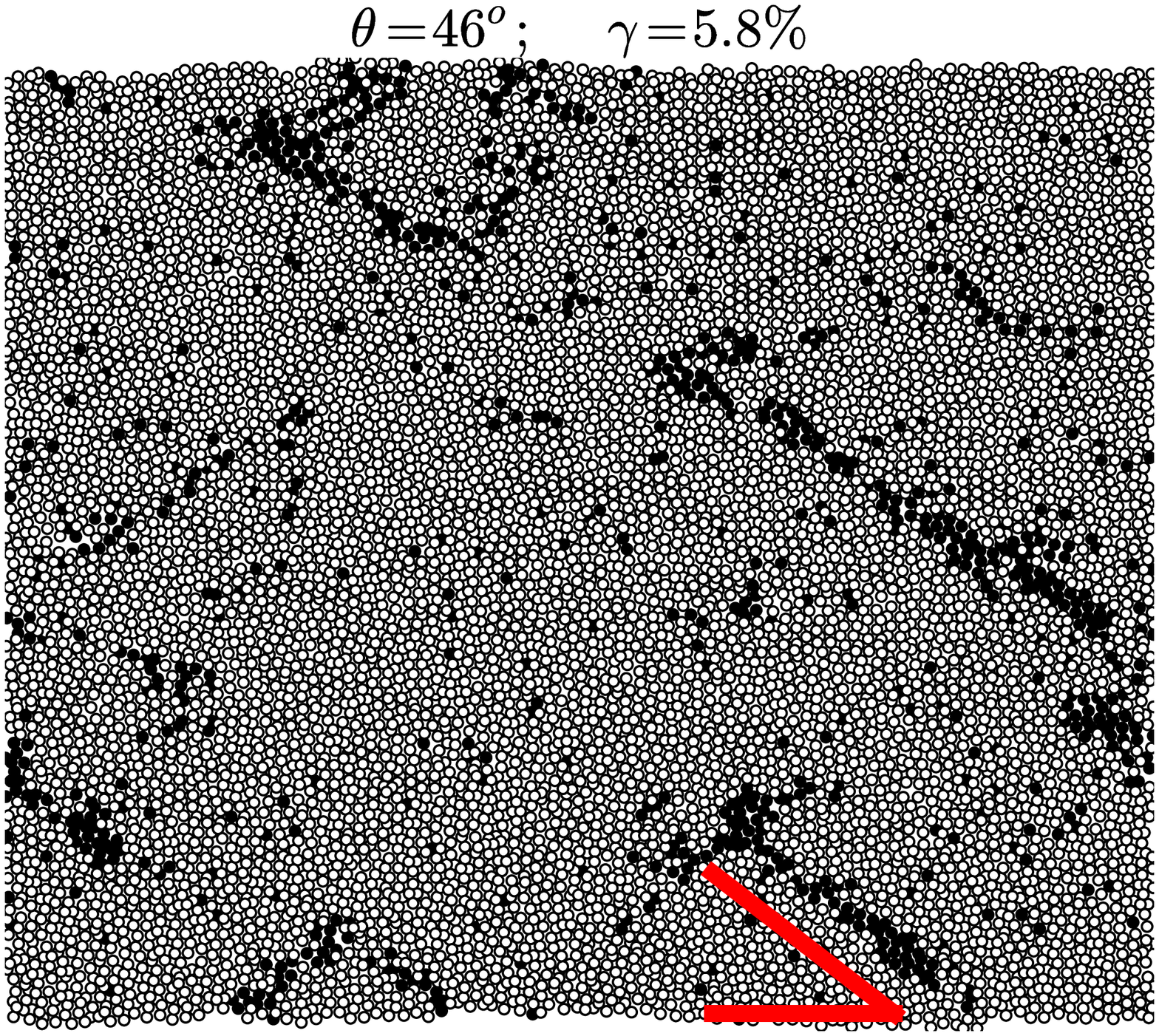}
\includegraphics[scale = 0.28]{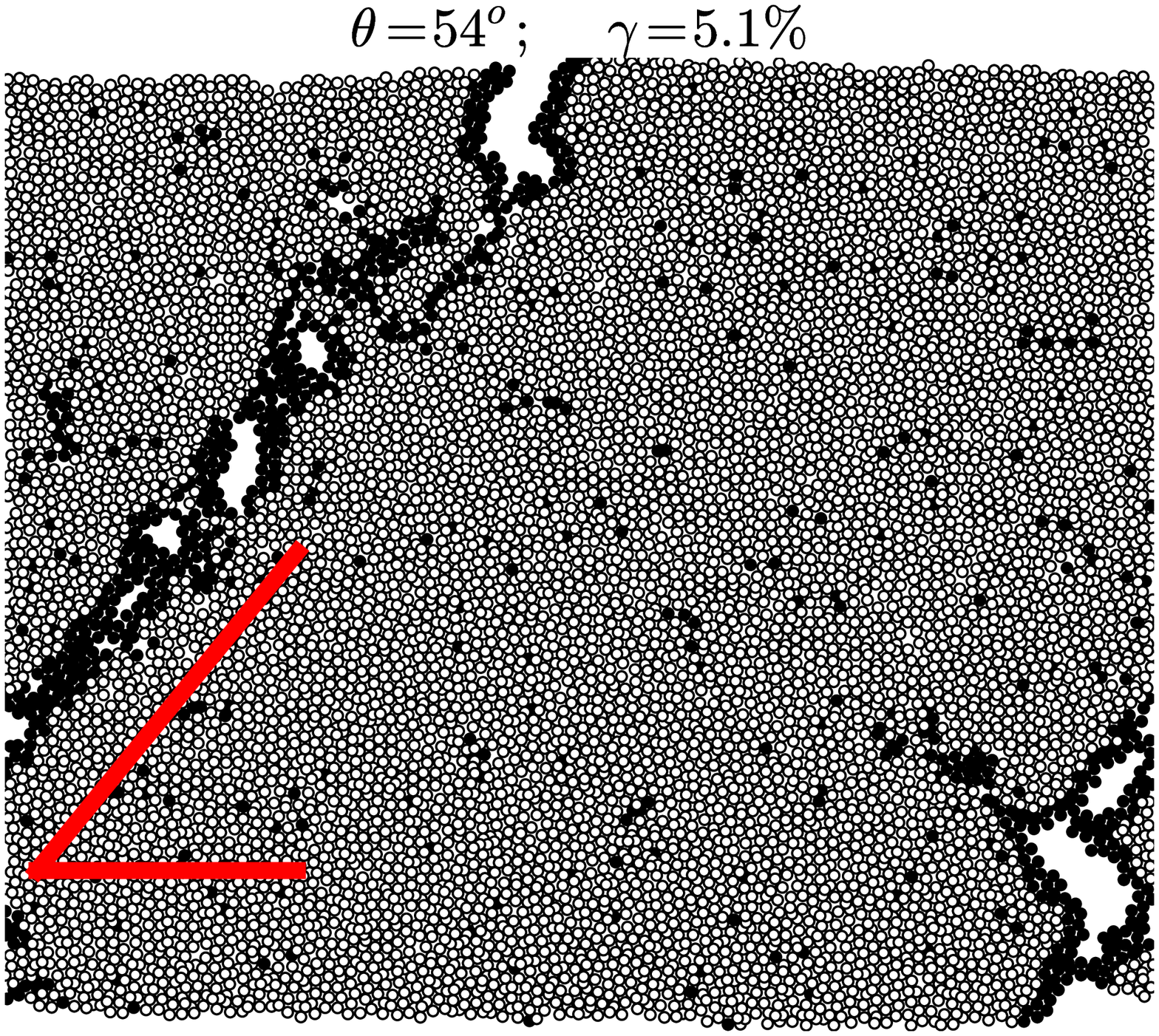}
\caption{Color Online: The shear band that occurs in a 2-dimensional amorphous solid upon uniaxial compression (upper panel) and extension (lower panel). The angle with respect to the principal stress component measured in the upper panel is 46$^o\pm 1^0$, whereas in the lower panel it is 54$^o\pm 1^0$. This Letter provides a microscopic theory of this asymmetry.}
\label{angles}
\end{figure}

The riddle that this Letter aims to solve is demonstrated in Fig.~\ref{angles}. Amorphous solids put under different loading conditions display shear bands that appear at different angles with respect to the principal stress axis \cite{11GWBN}. In past work phenomenological models were applied to this riddle \cite{75RR,08ZL,09ZL}, but a microscopic approach was lacking.
 In this Letter we provide a microscopic theory of this phenomenon in athermal systems that are strained under quasi-static conditions \cite{06ML,06TLB}. The Letter culminates with an analytic formula for the angle in terms of one parameter which is determined by the loading conditions. For simplicity and concreteness we will explain the
theory in 2-dimensions, the generalization to 3-dimension is available \cite{13DGPS}.

\begin{figure}
\includegraphics[scale = 0.28]{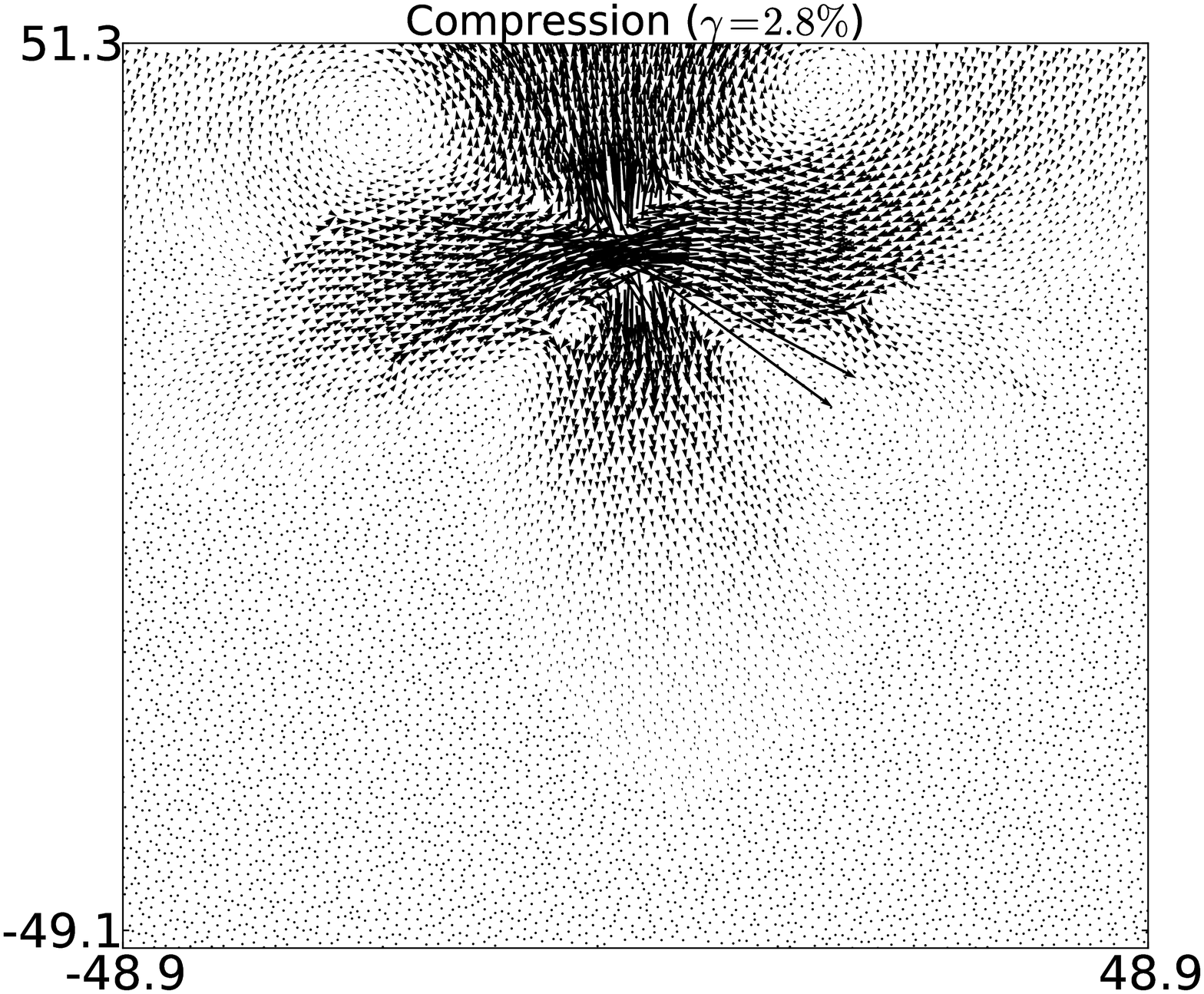}
\includegraphics[scale = 0.28]{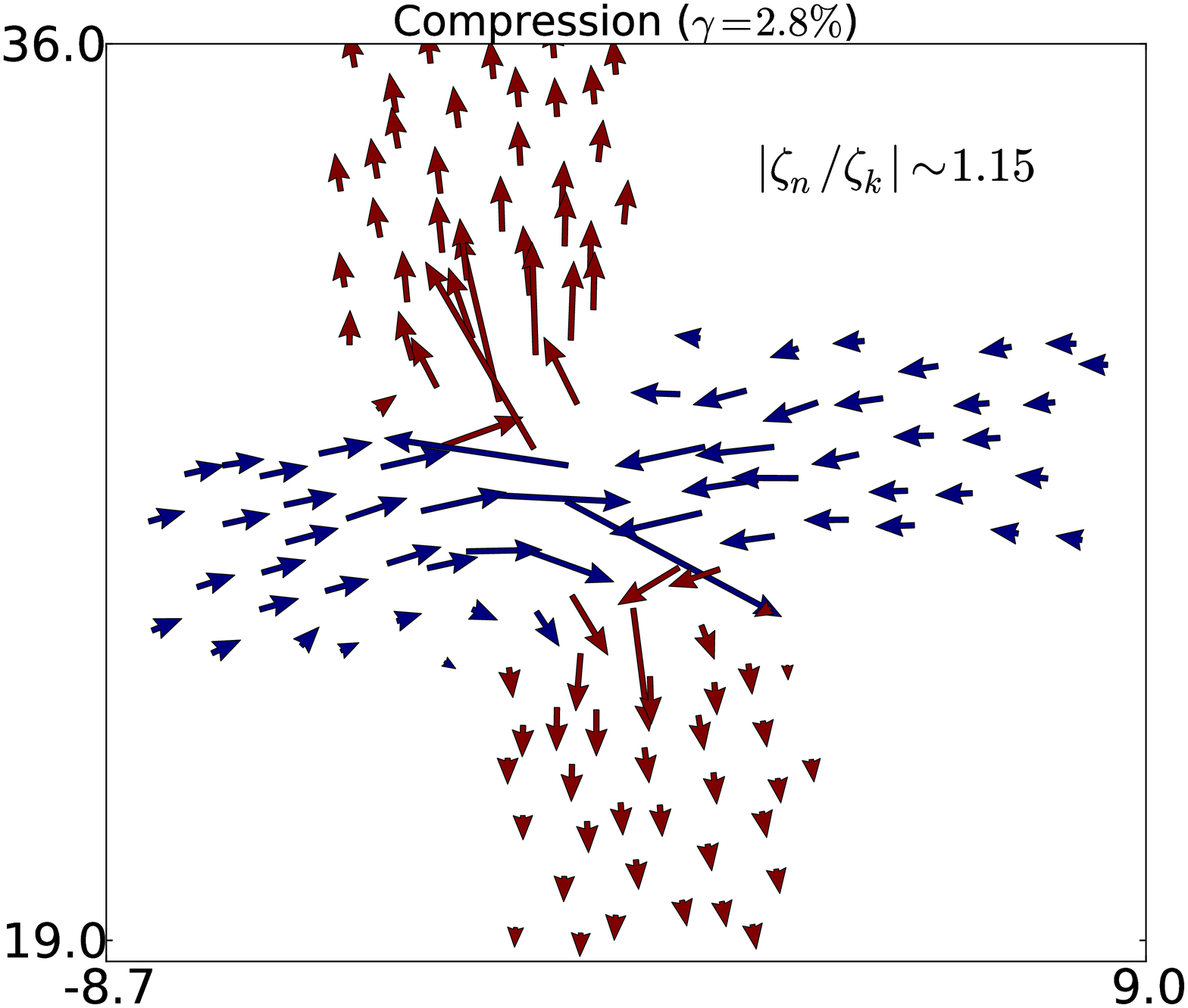}
\caption{Upper panel: A fundamental plastic event during an AQS uniaxial compression simulation of a 2-dimensional amorphous solid. Shown is the non-affine displacement field in the whole system.
Lower panel: a small window around the core of the event shown in the upper panel. For clarity we show only the
incoming and outgoing arrows.}
\label{eshcomp}
\end{figure}

Recently there had been rapid progress in understanding the nature of plastic instabilities in Athermal Quasistatic (AQS) conditions \cite{09LPa,09LPb,10KLLP,12DHP,13DHP}.
Consider a glass-forming system made of $N$ particles interacting via generic potentials (i.e. at least twice differentiable everywhere). The total energy can be written at $T=0$ in terms of the positions $\B r_1, \B r_2, \cdots \B r_N$ of these particles, $U = U(\B r_1,\B r_2 \cdots, \B r_N)$.  The Hessian matrix is defined as the
second derivative \cite{99ML}
\begin{equation}
H_{ij} \equiv \frac{\partial^2 U(\B r_1,\B r_2 \cdots, \B r_N)}{\partial \B r_i\partial\B r_j} \ .
\end{equation}
The Hessian is real and symmetric, and therefore can be diagonalized. Excluding Goldstone modes whose eigenvalues
are zero due to continuous symmetries, all the other eigenvalues are real and positive as long as the system
is mechanically stable. In equilibrium, without any mechanical loading, eigenfunctions associated with large eigenvalues are localized due to Anderson localization. But all the eigenfunctions associated with low eigenvalues (including all the excess modes that are typical to amorphous solids) are extended. At {\em low values of external loading} one observes ``fundamental plastic instabilities" when eigenvalues of some modes approach zero via a saddle
node bifurcation \cite{12DKP}. Simultaneously the associated eigenfunction localizes on a typical quadrupolar structure (in both 2 and 3 dimensions) that is identical to the non-affine displacement field associated with the plastic event. An
example of such a localized eigenfunction that is seen upon uniaxial compression is shown in Fig. \ref{eshcomp}.
At higher values of the external strain the nature of the plastic instability can change drastically. For well
quenched amorphous solids one finds that instead of single fundamental events there exists a bifurcation to the simultaneous appearance of a finite {\em density} of such events, (i.e. infinity of them in the thermodynamic limit), and that in two dimensions they organize along a line \cite{12DHP,13DHP}. In previous work the angle of this line with respect to the
principal stress axis was calculated analytically for external shear, with the result of $45^o$. We will explain now why external shear is special, and why the angle will change for other loading conditions.

A very important realization is that the fundamental plastic event can be excellently modeled analytically as
an Eshelby inclusion \cite{57Esh}. As is well known, the Eshelby construction in 2-dimensions starts with a circle taken out from an elastic medium, made into an ellipse that is then pushed back to the medium. This is the Eshelby inclusion. In terms of the two orthogonal directions $\hat{\B n}$ and $\hat{\B k}$ one writes the eigenstrain of the inclusion as
\begin{equation}
  \epsilon^{*}_{\alpha\beta} = \zeta_n\hat n_{\alpha}\hat n_{\beta} + \zeta_k\hat k_{\alpha}\hat k_{\beta} \ .
  \label{epsstar}
\end{equation}
The ``eigenvalues" $\zeta_n$ and $\zeta_k$ characterize the relative magnitude of the non-affine displacement
associated with the fundamental plastic instability. The direction $\hat n$ is that of the principal stress axis,
 and $\hat k$ the orthogonal one. The ratio $|\zeta_n/\zeta_k|$ depends, as shown below, on the loading
conditions. When area preserving prevails, $\zeta_n=-\zeta_k$ \cite{12DHP,13DHP}, but in general these numbers are independent, and
their ratio characterizes the loading condition. We argue below that this ratio determines uniquely the angle
of the shear band once there appears a density of such events with a prescribed ratio $|\zeta_n/\zeta_k|$.

Having the eigenstrain (\ref{epsstar}) one
asks what is the displacement field associated with such an inclusion. The calculation is shown in detail in the accompanying material \cite{supp}, with the final result
\begin{widetext}
\begin{align}
  \bm{u}^c(\bm{X}) &= \frac{(\zeta_n - \zeta_k)(\lambda+\mu)}{4(\lambda+2\mu)}\biggl(\frac{a^2}{r^2}\biggr) \biggl[2\frac{(\zeta_n + \zeta_k)}{(\zeta_n -
  \zeta_k)} \bm{X} + \biggl(\frac{2\mu}{\lambda+\mu} + \frac{a^2}{r^2}\biggr) \biggl(2(\hat{\bm{n}}\cdot\bm{X})\hat{\bm{n}} - \bm{X}\biggr) \nonumber\\
&+ 2\biggl(1- \frac{a^2}{r^2}\biggr)\biggl(2 (\hat{\bm{n}}\cdot\hat{\bm{r}})^2 - 1\biggr)\bm{X} \biggr]
  \label{uc-final}
\end{align}

\end{widetext}
with $\B X$ denoting an arbitrary Cartesian
point in the material, $r=|\B X|$,  $\B {\hat r}\equiv \B X/r$ and $a$ is a parameter known as the ``core size". The stress field induced by this displacement field is denoted below as $\sigma^{c}_{\alpha\beta}$.

As said above under simple shear one conserves area, and $\zeta_n=-\zeta_k$. Thus the
first term inside the square parenthesis in Eq. \ref{uc-final} vanishes, and in the two other terms $\zeta_n-\zeta_k =2\zeta_n\equiv \epsilon$, cf. \cite{13DHP}. We have seen in Ref. \cite{13DHP}, and we will see below, that when such a condition applies, the angle of the shear band is precisely 45$^o$ with respect to the principal stress axis. But for the more general
loading conditions in which $\zeta_n\ne -\zeta_k$ the angle is different, see Eq. (\ref{final}) below.

When the amorphous solid is well quenched, the nature of the plastic instability
changes when the external strain becomes sufficiently high; a density of inclusions of the form (\ref{uc-final})
appear simultaneously. To find their geometrical organization we need to compute the energy of such $\C N$
inclusion in a system of total volume $V$ and minimize it with respect to their orientation and geometry. The total energy of such $\C N$ inclusion has four contributions:
\begin{equation}
   E = E_{mat} + E_{\infty} + E_{esh} + E_{inc}
   \label{E}
 \end{equation}
 with each component of energy defined as
  \begin{align}
   E_{mat} &= \frac{1}{2}\sigma^{\infty}_{\alpha\beta} \epsilon^{\infty}_{\beta\alpha} V \label{Emat}  \\
   E_{\infty} &= -\frac{1}{2}\sigma^{\infty}_{\alpha\beta} \sum^{N}_{i=1}\epsilon^{*,i}_{\beta\alpha}V^i_0 \label{Einf}\\
   E_{esh} &= \frac{1}{2}\sum_{i=1}^{N} (\sigma^{*,i}_{\alpha\beta} - \sigma^{c,i}_{\alpha\beta})\epsilon^{*,i}_{\beta\alpha} V^i_0 \label{Eesh}\\
   E_{inc} &= -\frac{1}{2} \sum^{N}_{i=1}\epsilon^{*,i}_{\beta\alpha} V^i_0 \sum_{j\neq i} \sigma^{c,j}_{\alpha\beta}(r^{ij}) \ . \label{Einc}
 \end{align}
 Here the eigen-strain $\epsilon^{*,i}_{\alpha\beta}$ and volume $V^i_0$ associated with any $i^{th}$ Eshelby inclusion are given as
 \begin{align}
   V^i_0 &= \pi a^2 \nonumber\\
   \epsilon^{*,i}_{\alpha\beta} &= \frac{(\zeta_n - \zeta_k)}{2}(2 \hat{n}^i_\alpha \hat{n}^i_\beta - \delta_{\alpha\beta}) + \frac{(\zeta_n +
  \zeta_k)}{2}\delta_{\alpha\beta}
  \label{Eshelby-property}
 \end{align}
 The reader should recognize that the first contribution, (\ref{Emat}), is the energy of the elastic matrix and the second ((\ref{Einf})) is
 the energy due to the interaction of the elastic matrix with the inclusions. The third, (\ref{Eesh}), is the self-energy
 of all the inclusions, and the last, (\ref{Einc}), is the interaction energy of different inclusions.

 Also for a 2D material being loaded under uni-axial strain with free boundaries along $\hat{y}$, we can write the form of the global stress tensor as
 \begin{equation}
   \sigma^{\infty} = \left( \begin{array}{cc}
     \sigma^\infty_{xx} & 0 \\
			0 & 0 \\
   \end{array} \right) \ .
   \label{global-stress-tensor}
 \end{equation}
 By Hooke's law, we get the expression for applied global stress tensor
 using the Lam\'e coefficients $\lambda$ and $\mu$:
 \begin{equation}
   \sigma^{\infty}_{\alpha\beta} = \lambda \epsilon^{\infty}_{\eta\eta}\delta_{\alpha\beta} + 2\mu \epsilon^{\infty}_{\alpha\beta} \ . \label{glostress1}
 \end{equation}
  Taking trace of Eq. (\ref{glostress1}), we find
 \begin{equation}
   \epsilon^\infty_{\eta\eta} = \frac{1}{2(\lambda+\mu)}\sigma^\infty_{\eta\eta} = \frac{1}{2(\lambda+\mu)}\sigma^\infty_{xx}
   \label{glostress2}
 \end{equation}

 Plugging Eq. (\ref{glostress2}) in Eq. (\ref{glostress1}), we find
 \begin{equation}
   \sigma^\infty_{xx} = \frac{4\mu(\lambda+\mu)\gamma}{\lambda + 2\mu}
   \label{global-stress-3}
 \end{equation}

 where $\gamma$ is the external strain.

Clearly, the first contribution Eq. (\ref{Emat}) is independent of the distribution of inclusions and will not
affect the state of minimal energy. The second term Eq. (\ref{Einf}) is the only one proportional to the external
strain $\gamma$, and therefore for sufficiently large $\gamma$ it needs to be minimized first. Besides $\gamma$,
this term depends only on the orientation of each inclusion, i.e. on the angle $\phi = \cos^{-1}(n^i_x) = \sin^{-1}(n^i_y)$. Minimizing the term with respect to this angle we find that the minimum is obtained when
$\phi=0$ or $\pi/2$. In other words, each inclusion is oriented with one axis parallel and the other
orthogonal to the uniaxial direction. Plugging this information into Eqs. (\ref{Eesh}) and (\ref{Einc}) simplifies
them considerably, and see the supplementary material for full details \cite{supp}. The organization and orientation of the density of inclusions is determined by minimizing $E_{\rm inc}$ of Eq. (\ref{Einc}). The minimum energy
is obtained when all the inclusions align on a line that is at an angle $\theta$ with respect to the uniaxial direction \cite{supp}. The final prediction for this angle is
\begin{equation}
\cos^2\theta =\frac{1}{2} - \frac{\zeta_n+\zeta_k}{4(\zeta_n-\zeta_k)} \ ,
\end{equation}
or
\begin{equation}
 \theta = \cos^{-1}\sqrt{\frac{1}{2} - \frac{\zeta_n + \zeta_k}{4(\zeta_n - \zeta_k)}}\ . \label{final}
\end{equation}

\begin{figure}
\includegraphics[scale = 0.28]{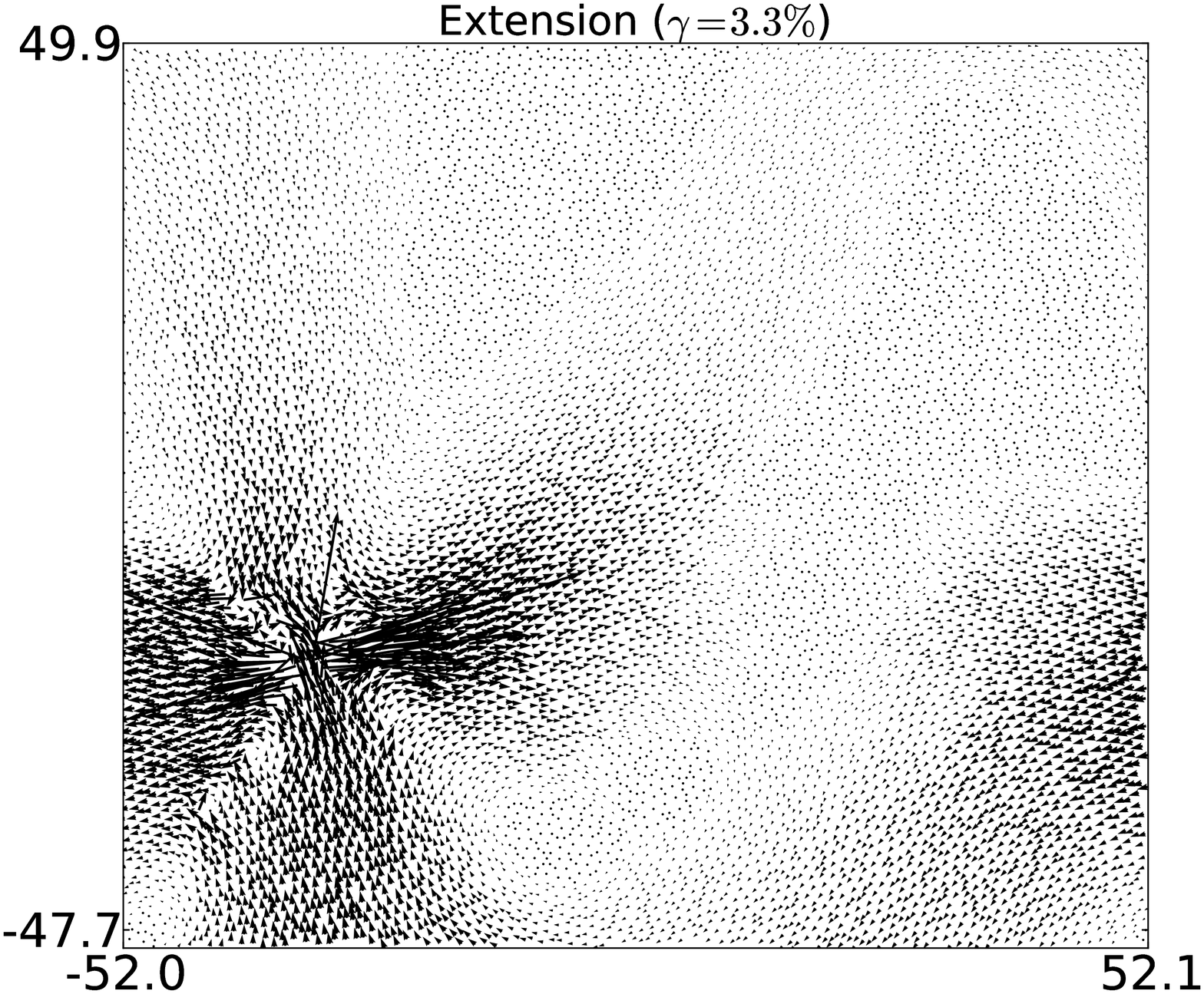}
\includegraphics[scale = 0.28]{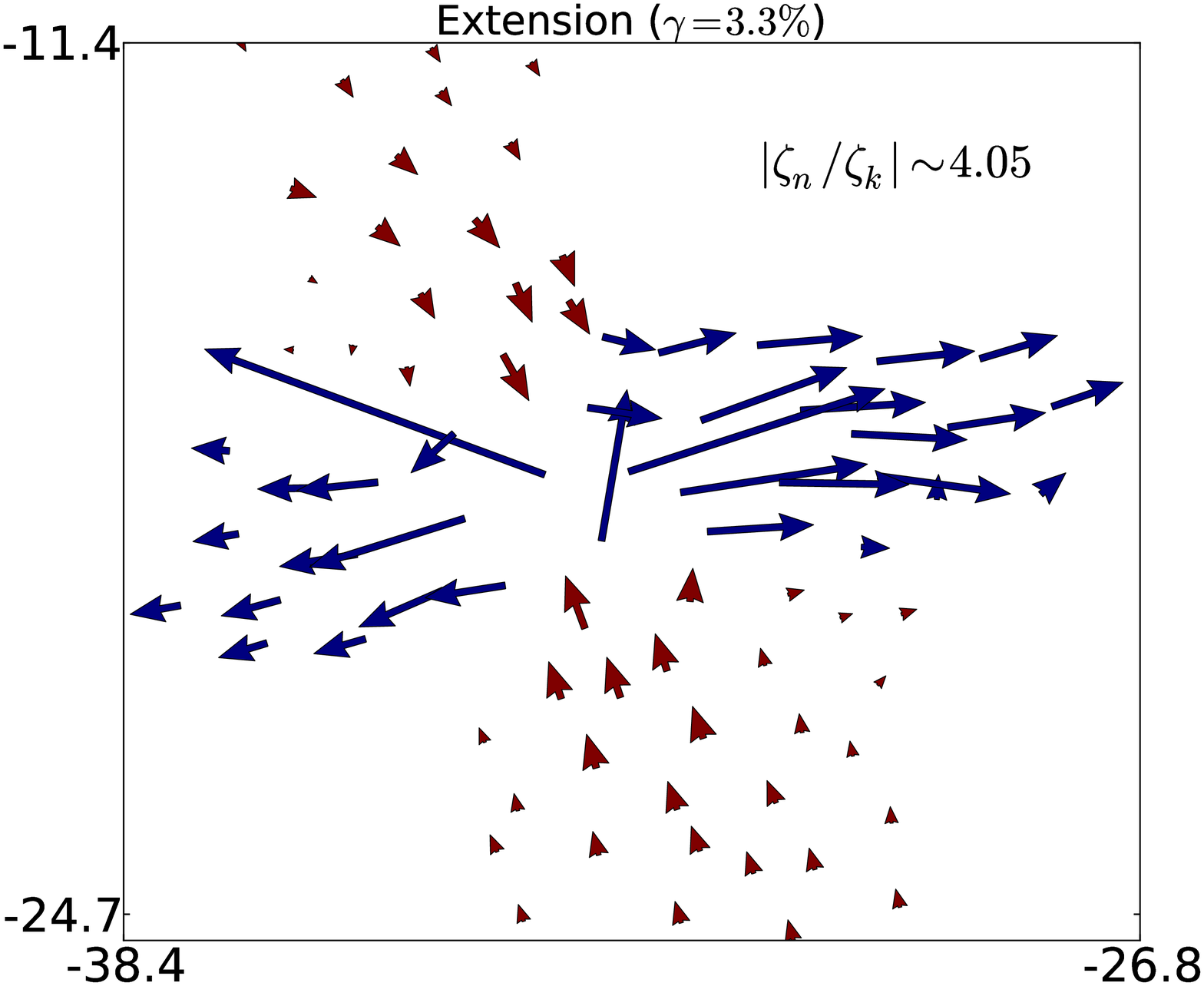}
\caption{Upper panel: A fundamental plastic event during an AQS uniaxial extension simulation of a 2-dimensional amorphous solid. Shown is the non-affine displacement field in the whole system.
Lower panel: a small window around the core of the event shown in the upper panel. For clarity we show only the
incoming and outgoing arrows.}
\label{eshextension}
\end{figure}

One learns immediately from Eq. (\ref{final}) that the area preserving case, $\zeta_n=-\zeta_k$, leads
to an angle of precisely $45^o$. Any other loading condition would result in a different angle. The two extreme cases
occur for $|\zeta_n/\zeta_k|\to 0$ and $|\zeta_n/\zeta_k|\to \infty$. The first case would realize an angle
of $30^o$ and the second an angle of $60^o$. All experimental realizations should fall between these two
extreme universal limits without exception. Indeed, examining the wide range of angles reported in Ref. \cite{11GWBN}
we see that all the data falls within our theoretical limits. We return now to our simulations shown in Fig. \ref{angles} to rationalize the angles observed.

In order to understand the angle seen in a particular experiment one needs to figure how the values of
$\zeta_n$ and $\zeta_k$ are determined by the loading conditions. For example in uniaxial extension the outgoing direction may become highly dominant compared to the incoming one, leading to a high ratio of $|\zeta_n/\zeta_k|$. This is not the case for compression, as one can see from Fig. \ref{eshcomp}. We expect such asymmetry to appear in
any generic potential due to the steep rise of the repulsive part vs the moderate attractive tail. The ratio of eigenvalues is obtained from the lower panel by averaging the length of the incoming and outgoing vectors around the core. We find that the ratio of the average length provides a fair estimate of $|\zeta_n/\zeta_k|$, in this case $|\zeta_n/\zeta_k|\approx 1.15$. Using this value in Eq. (\ref{final}) we find an angle $\theta\approx 46^o$ in perfect agreement with the observed angle in Fig. \ref{angles} upper panel. The exercise is repeated for uniaxial extension, see Fig. \ref{eshextension}. Obviously, in the extension case the outgoing arrows are much longer than the incoming ones. Averaging the length
of the outgoing and incoming vectors and taking the ratio we find for the present case $|\zeta_n/\zeta_k|\approx
4.05$. Plugging this value into Eq. (\ref{final}) we get in this case $\theta\approx 54^o$, again in perfect accord
with the observed angle in Fig. \ref{angles} lower panel.

To summarize, we note that the atomistic theory is based in the understanding that shear localization occurs
due to a plastic instability in which a density of Eshelby-like quadrupolar structures organize on a line.
Each individual quadrupolar structure is sensitive to the loading conditions, parameterized in the 2D case
by the ratio of ``eigenvalues" $\zeta_n/\zeta_k$. At zero temperature and under quasi-static conditions
one is well justified to find the preferred geometry by energy minimization. This leads to the prediction that
all the quadrupoles have the same orientation (for sufficiently high external strain) and that they organize
on a line at an angle $\theta$ with respect to the principal stress axis. The angle $\theta$ is limited in this theory to fall between $30^o$ and $60^o$, with $45^o$ being special to the area preserving situation $\zeta_n/\zeta_k=-1$. This range of allowed angles appears to be well in accord with published experimental results.

We have considered in detail the orientation of the shear bands under uniaxial compression and extension and
demonstrated the sensitivity of the ratio of `eigenvalues' to these different loading conditions. Finally,
we showed that the theoretical prediction Eq. (\ref{final}) is in excellent agreement with the observed angles
in these loading conditions.

The extension of the theory to finite temperatures and strain rates is beyond the scope of this Letter,
but see Ref. \cite{13DJHP} for a possible direction to go. Also, in the follow up paper to this Letter
we will present the theory for the asymmetry in the yield-strain as a function of $\zeta_n$ and $\zeta_k$. Stay tuned.

{\bf Acknowledgements}: this work was supported by the Israel Science Foundation, the German-Israeli Foundation and
by the ERC under the STANPAS ``ideas" grant.


\begin{thebibliography}{99}

\bibitem{11GWBN}
For extensive experimental evience see Y.F. Gao, L. Wang, H. Bei and T.G. Nieh, Acta Materiala,  {\bf 59}, 4159 (2011),
and references therein.

\bibitem{75RR}
J.W. Rudnicki and J. R. Rice, J. Mech. Phys. Solids, {\bf 23}, 371 (1975).

\bibitem{08ZL}
M. Zhao and M. Li, Appl. Phys. Lett {\bf 93}, 241906 (2008).

\bibitem{09ZL}
M. Zhao and M. Li, J. Mater. Res. {\bf 24}, 2688 (2008).


\bibitem{06ML}
C.E. Maloney and A. Lema\^itre, Phys. Rev. E {\bf 74}, 016118 (2006).

 \bibitem{06TLB}
A. Tanguy, F. Leonforte and J.L Barrat, Eur. Phys. J. {\bf E20}, 355-364 (2006).

\bibitem{13DGPS}
R. Dasgupta, O. Gendelman, I. Procaccia and C. Shor ``Shear localization in 3-Dimensional Amorphous Solids",
to be submitted to PRE.

\bibitem{09LPa}
E. Lerner and I Procaccia, Phys Rev E, {\bf 79},066109 (2009).

\bibitem{09LPb}
E. Lerner and I. Procaccia, Phys, Rev. E, {\bf 80}, 026128 (2009).

\bibitem{10KLLP}
S. Karmakar, A. Lema\^itre, E. Lerner and I. Procaccia, Phys. Rev. Lett. {\bf 104}, 215502 (2010).

\bibitem{12DHP}
R. Dasgupta, H. G. E. Hentschel and I. Procaccia,  Phys.Rev. Lett., {\bf 109} 255502 (2012).

\bibitem{13DHP}
R. Dasgupta, H. G. E. Hentschel and I. Procaccia, Phys. Rev. E {\bf 87}, 022810 (2013).

\bibitem{99ML}
D. L. Malandro and D. J. Lacks, J. Chem. Phys. {\bf 110}, 4593 (1999).

\bibitem{12DKP}
R. Dasgupta, S. Karmakar and I. Procaccia, Phys. Rev. Lett. {\bf 108}, 075701 (2012).

\bibitem{57Esh}
  J. D. Eshelby, Proc. R. Soc. Lond. A {\bf 241}, 376 (1957); {\bf 252}, 561 (1959).

\bibitem{supp}
http://www.weizmann.ac.il/chemphys/cfprocac/\\publ.html (papers online \#191).

\bibitem{13DJHP}
R. Dasgupta, Ashwin J., H.G.E.Hentschel and I. Procaccia,  Phys. Rev. B {\bf 87}, 020101(R) (2013).

\end{thebibliography}
\end{document}